\documentclass[11pt]{article}

\usepackage{amsmath,amssymb,amsthm}
\usepackage{ulem}
\usepackage{tikz}
\usepackage{nicefrac}
\usepackage{slashed}
\usepackage{empheq}
\usepackage{xcolor}
\usepackage[shortlabels]{enumitem}

\usepackage[left=1in, right=1in, top=1in, bottom=1in]{geometry}
\usepackage{cancel}

\numberwithin{equation}{section}
\theoremstyle{remark}

\newcommand{\be}{\begin{equation}}
\newcommand{\ee}{\end{equation}}

\definecolor{darkgreen}{rgb}{0.0,0.5,0.0}
\definecolor{darkblue}{rgb}{0.0,0.0,0.3}
\definecolor{nicosred}{rgb}{0.65,0.1,0.1}
\definecolor{light-gray}{gray}{0.7}

\usepackage{todonotes} 

\usepackage{hyperref}

\title{To trace or not to trace: analytical insights from network-based contact-tracing models}

\author{Giulia de Meijere$^{1,*}$, Andrea Pugliese$^{2}$, Gerardo I\~niguez$^{1,3}$, P\'eter L. Simon$^4$ and Istv\'an Z. Kiss$^{5,6,*}$}
\date{\scriptsize{
$^1$ Tampere Complexity Lab, Data Science Research Centre, Tampere University, FI-33720 Tampere, Finland\\
$^2$ Department of Mathematics, University of Trento, Via Sommarive 14, 38123 Trento, Italy\\
$^3$ Centro de Ciencias de la Complejidad, Universidad Nacional Autonóma de México, 04510 Ciudad de México, Mexico\\
$^4$ Institute of Mathematics, E\"otv\"os Lor\'and University, Budapest, and National Laboratory for Health Security, Hungary\\
$^5$ Network Science Institute, Northeastern University London, London E1W 1LP, UK\\
$^6$ Department of Mathematics, Northeastern University, Boston, MA 02115, USA\\}
$^*$ Corresponding author email: giulia.demeijere@tuni.fi, istvan.kiss@nulondon.ac.uk \\
	  \vspace{1cm}\today
	  }

\begin{document}
\maketitle	

\begin{abstract} 
Contact tracing is one of the most important control measures deployed during epidemics and pandemics, relying on the identification of contacts of known infected individuals. A network perspective is therefore essential for its accurate modelling. Pairwise models have been used extensively to study contact tracing, but their analysis typically depends on a decoupling assumption—most commonly that contact tracing operates on a much faster timescale than disease transmission. Furthermore, contact tracing models often assume that each infected individual becomes a contact tracing-triggering node, which is unrealistic given partial compliance to treatment in practice.
In this paper, we relax both of these restrictive assumptions and provide a full analytical characterisation of the epidemic threshold in the pairwise mean-field model. Our analysis uses a fast-variables approach that captures the rapid early stabilisation of key network quantities. In addition, inspired by mechanisms from social adoption dynamics, we introduce triplewise contact tracing in which an infected individual can be traced not only through direct contact with a single tracing-triggering neighbor (pairwise tracing), but also indirectly when connected to two tracing-triggering nodes simultaneously.
For pure pairwise and pure triplewise contact tracing, we derive analytical expressions for critical contact tracing thresholds and demonstrate that when many infected individuals bypass treatment, the epidemic can become uncontrollable. When both contact tracing mechanisms operate together, we map out their combined contribution and relative impact on epidemic control. This unified framework yields rigorous and tractable threshold conditions for contact tracing dynamics on networks, extending the applicability of pairwise models beyond the fast-tracing regime and providing new insight into the interplay between disease progression, partial treatment compliance, and higher-order tracing processes.
\end{abstract}


\section{Introduction}

The silent spread of infectious diseases poses a fundamental challenge to epidemic control, particularly when infections exhibit pre-symptomatic transmission, mild symptoms, or asymptomatic progression. Contact tracing---the systematic identification and monitoring of individuals who have been exposed to infected cases---has emerged as a critical non-pharmaceutical intervention that leverages network structure rather than symptomatology to identify at-risk individuals \cite{rutherford1988contact}. The theoretical foundation for understanding contact tracing efficacy lies at the intersection of network science and mathematical epidemiology, where the explicit consideration of contact patterns becomes essential for accurate predictions \cite{keeling2005networks}.

The mathematical modeling of contact tracing has evolved considerably since the seminal work of Eames and Keeling \cite{eames2003contact}, who introduced the Suceptible-Infected-Tracing-Recovered (SITR) framework that explicitly accounts for traced individuals. Their analysis, however, relied on a critical simplifying assumption: that contact tracing operates on a much faster timescale than disease transmission. This assumption, while mathematically convenient, may not hold in practice where tracing delays, limited resources, and imperfect compliance affect the speed and completeness of tracing efforts \cite{hellewell2020feasibility}. Recent empirical evidence from COVID-19 has highlighted that the effectiveness of contact tracing is highly sensitive to delays and adherence rates \cite{ferretti2020quantifying,davis2021contact}, suggesting that relaxing the fast-tracing assumption is crucial for realistic assessment of control strategies.

Network structure fundamentally shapes both disease transmission and the efficacy of contact tracing \cite{pastor-satorras_epidemic_2015}. While mean-field models provide analytical tractability, they fail to capture the heterogeneity and clustering present in real contact networks \cite{kiss2017mathematics}. Pairwise approximation models offer a middle ground, incorporating local network structure while maintaining analytical accessibility \cite{barnard2019epidemic}. These models explicitly track pairs of individuals in different states, allowing for the representation of correlations that develop during epidemic spread. The fast-variables approach introduced by Barnard et al. \cite{barnard2019epidemic} demonstrated that network quantities such as the average number of susceptible neighbors per infected individual reach quasi-equilibrium rapidly, providing a novel pathway to derive epidemic thresholds without requiring the problematic linearization around disease-free steady-state.

The mechanisms of contact tracing itself have also evolved beyond simple forward tracing. Bidirectional tracing, which identifies both contacts and sources of infection, has been shown to dramatically improve control efficacy \cite{bradshaw2021bidirectional}. Similarly, backward tracing---prioritizing the identification of infection sources---can be particularly effective in networks with heterogeneous transmission \cite{kojaku2021effectiveness}. These enhanced tracing strategies suggest that the classical pairwise tracing model may be insufficient to capture the full spectrum of tracing mechanisms available to public health authorities.

Recent theoretical advances have emphasized the importance of higher-order network structures in epidemic dynamics. House et al. \cite{house2009motif} developed a motif-based approach that systematically incorporates triangles and other small subgraphs, demonstrating that clustering and other higher-order structures can significantly alter epidemic thresholds. This motivates the consideration of triplewise interactions in contact tracing, where an individual might be traced through multiple pathways or require multiple exposures to tracing signals before taking action---a mechanism analogous to complex contagion in social networks.

The timing and effectiveness of contact tracing are constrained by fundamental epidemiological parameters~\cite{mettler2021importance}. Fraser et al. \cite{fraser2004factors} identified key factors determining outbreak controllability, emphasizing the critical role of the proportion of transmission occurring before symptom onset. For diseases with substantial pre-symptomatic transmission, the window for effective contact tracing narrows considerably, making rapid and comprehensive tracing essential. This underscores the importance of developing models that can accurately capture the interplay between tracing speed, coverage, and disease natural history.

The comprehensive review by M\"uller and Kretzschmar \cite{muller2021contact} highlights the diversity of contact tracing models and their varying predictions about efficacy, emphasizing the need for careful consideration of model assumptions. Network-based approaches have proven particularly valuable, as demonstrated by Miller \cite{miller2011note} and extended in the comprehensive treatment by Kiss, Miller, and Simon \cite{kiss2017mathematics}. The challenge of selecting appropriate model structures for emerging epidemics has been systematically addressed by Pellis et al. \cite{pellis2015systematic} and Ball et al. \cite{ball2015seven}, who emphasize the trade-offs between model complexity and analytical tractability.

In this paper, we extend the analytical framework for contact tracing on networks by addressing two key limitations of existing models. First, we relax the assumption of infinitely fast contact tracing, allowing for realistic timescales where tracing and transmission operate concurrently. Second, we introduce a novel triplewise contact tracing mechanism, where individuals require exposure to multiple traced neighbors before taking action. This mechanism captures scenarios where social reinforcement or institutional thresholds govern behavioral responses to tracing notifications. Using the fast-variables approach, we derive explicit threshold conditions for epidemic control under both pairwise and triplewise tracing, providing analytical insights that remain valid across the full parameter space. Our results reveal that contact tracing effectiveness depends critically on network structure, with the required tracing effort scaling non-linearly with network degree and clustering. These findings provide a rigorous foundation for designing contact tracing strategies that account for realistic operational constraints and behavioral responses.

\section{Model}
\label{sec:models}

\begin{figure}
    \centering
    \includegraphics[width=9cm]{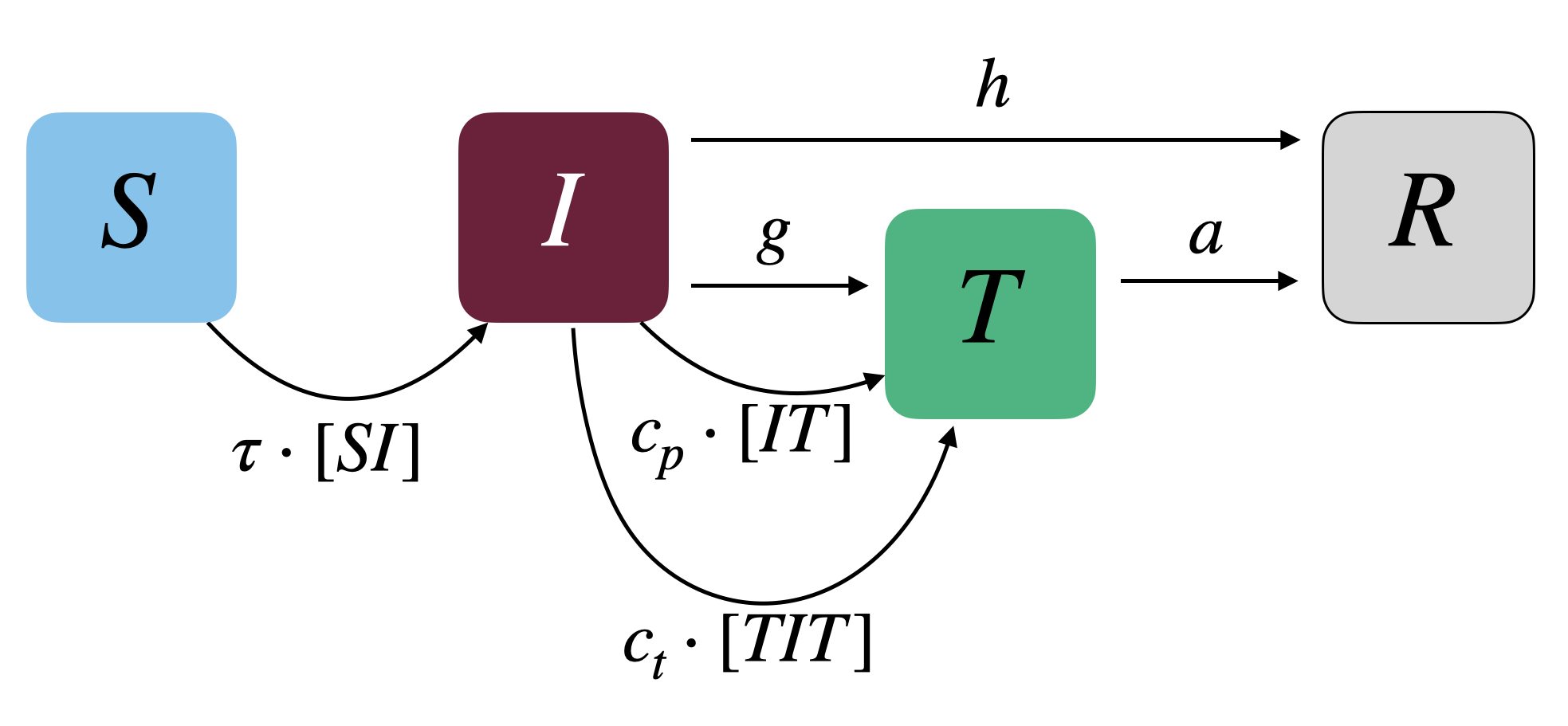}
    \caption{\textbf{Schematic description of the epidemic model.} Individuals exist in four states: susceptible $S$, infected $I$, treated (tracing) $T$, and recovered $R$. A susceptible individual becomes infected at rate $\tau$, conditional on an $[SI]$ contact. An infected individual either recovers naturally at rate $h$ or seeks treatment at rate $g$, thereby moving to the treated state and initiating contact tracing. Treated individuals exit $T$ at rate $a$ and move to $R$. In addition, infected individuals move to $T$ through contact tracing at rate $c_p$ conditional on an $[IT]$ contact, and at rate $c_t$ conditional on a $[TIT]$ triple.}
    \label{fig:schematic_model}
\end{figure}

We propose an extension of the `SITR' (Suceptible-Infected-Tracing-Recovered) epidemic model for contact tracing (CT) introduced in~\cite{eames2003contact}.
The model of~\cite{eames2003contact} builds on the standard SIR (Susceptible–Infected–Recovered) epidemic framework \cite{kermack1927contribution}. New infections occur at rate $\tau$, conditional on contact between a susceptible individual $S$ and an infected, infectious individual $I$. Infected individuals leave the infectious state at rate $g$ and enter a treated (tracing) state $T$, in which they are no longer contagious. This state represents, for example, individuals who recover prematurely due to treatment or who adopt interventions that fully prevent onward transmission, such as effective mask use. Importantly, treated individuals may also influence their infected neighbors to seek treatment. This contact tracing mechanism induces additional transitions from $I$ to $T$ at rate $c_p$, conditional on pairwise interaction between treated and infected individuals. Finally, treated individuals cease tracing activity at rate $a$ and move to the recovered state $R$, which is absorbing.

In the present work, we propose two extensions of this skeleton model. First, we allow a partial compliance of the infected population to treatment: infected individuals may recover following the natural course of infection, without ever seeking treatment. We call this the natural recovery rate $h$. Second, we allow for a higher-order version of contact tracing. Indeed, when it comes to social influence, multiple exposures might be required to trigger a transition. Here, we tackle this by requiring the exposure of an infected individual in $I$ to two treated individuals in $T$ for a transition to occur from $I$ to $T$, with new rate $c_t$. Note that this triplewise implementation of contact tracing is in general not equivalent to the standard absolute threshold model with threshold $\theta = 2$ \cite{starnini_opinion_2025}. Indeed, in our model, infection is only possible, not guaranteed, once the threshold has been reached. A schematic representation of the full compartmental model for CT is provided in Fig.~\ref{fig:schematic_model}.

To capture correlations arising from contact tracing, we formulate the model within the pairwise approximation framework~\cite{kiss2017mathematics}. In addition to tracking the expected number of individuals in each epidemiological state, we explicitly follow the dynamics of connected pairs of individuals. Let $[A]$ denote the expected number of nodes in state $A\in\{S,I,T,R\}$, and $[AB]$ the expected number of edges connecting nodes in states $A$ and $B$, irrespective of order. The evolution of node- and pair-level quantities is derived by accounting for all events that create or destroy nodes or edges of a given type. 
As tracing mechanisms depend explicitly on local network structure~\cite{eames2003contact}, higher-order motifs naturally arise in the description: triples of the form $[ABC]$ represent connected paths of length two, while four-node motifs $[AEBC]$ correspond to 3-stars centered at node $E$. The network motifs encode the local configurations required for pairwise and triplewise tracing events and appear explicitly in the resulting system of equations, as shown below:
\begin{align}
\begin{cases}
{[\dot{S}] } & = -\tau [S I]\\
{[\dot{I}] } & = \tau [S I] - (g + h) [I] - c_p [I T] - c_t [TIT] \\
{[\dot{T}] } & = g [I] + c_p [I T] - a [T] + c_t [TIT] \\
{[\dot{R}] } & = a [T] + h[I] \\
{[\dot{S S}] } & = -2 \tau [S S I] \\
{[\dot{S I}]} & = \tau([S S I] - [I S I] - [S I]) - c_p [S I T] - (g+h) [SI] - c_t [S I T T]\\
{[\dot{S T}]} & = -\tau [IST] + g[SI] + c_p[SIT] - a[ST] + c_t[SITT]\\
{[\dot{I I}] } & = 2 \tau ([S I]+[I S I])-2 (g+h)[I I] - 2c_p[I I T] - 2 c_t [I I T T]\\
{[\dot{I T}] } & = \tau[I S T] - a[I T] + c_p([I I T] - [T I T] - [I T]) + g [II] - (g + h) [IT] + c_t [I I T T] \\ 
& \quad - c_t ([T I T] + [T I T T]),\\
{[\dot{T T}]} & = 2 g [I T] + 2 c_p ([IT] + [T I T]) + 2 c_t ([TIT] + [T I TT]) - 2 a [T T].
\end{cases}
\end{align}
The terms proportional to $\tau$ correspond to infection events along $S-I$ edges, as in the standard pairwise SIR model. The terms proportional to $g$ and $h$ describe spontaneous transitions $I \to T$ (self-reporting or background detection) and $I \to R$, respectively, while $a$ governs removal of treated individuals. The terms proportional to $c_p$ capture pairwise contact tracing: these involve motifs such as $[IT]$, $[SIT]$, and $[IIT]$, reflecting tracing triggered by a treated neighbor. The terms proportional to $c_t$ represent triplewise tracing mechanisms and depend on higher-order motifs such as $[TIT]$ and $[IITT]$, encoding configurations in which tracing is activated through the presence of not one but two contacts in state $T$.

The model forms an open hierarchy, which may subsequently be closed using standard moment-closure approximations. We consider the following standard closure for the triple \cite{keeling2005networks}:
\begin{equation}
    [ABC] = \frac{(n-1)}{n} \frac{[AB] [BC]}{[B]},
\end{equation}
where $n$ is the average degree.
For the 4-motif, assuming a star structure and based on \cite{house2009motif}, the closure reads:
\begin{align}
    [A E B C] &= [*] \frac{[-]^3}{[<]^3[.]}\frac{[AEB][AEC] [BEC]}{[AE][BE][CE]} \cdot [E] \notag\\
    &= \frac{(n-1)(n-2)}{n^2} \frac{[AE] [BE] [CE]}{[E]^2},
\end{align}
where the number of nodes is $[\cdot]=N$, the number of edges is $[-] = n N$, the number of triples is $[<] = N n (n-1)$, and the number of 3-stars is $[*] = N n (n-1)(n-2)$.

Using these closures, the dynamics becomes:
\begin{align}
\label{eq:system}
\begin{cases}
{[\dot{S}] } & = -\tau [S I]\\
{[\dot{I}] } & = \tau [S I] - [I] - c_p [I T] - c_t \frac{n-1}{n} \frac{[IT]^2}{[I]} \\
{[\dot{T}] } & = q [I] + c_p [I T] - a [T] + c_t \frac{n-1}{n} \frac{[IT]^2}{[I]} \\
{[\dot{R}] } & = a [T] + (1-q) [I] \\
{[\dot{S S}] } & = -2 \tau \frac{n-1}{n} \frac{[SS] [SI]}{[S]} \\
{[\dot{S I}]} & = \tau(\frac{n-1}{n} \frac{[SS] [SI]}{[S]} - \frac{n-1}{n} \frac{[SI]^2}{[S]} - [S I]) - c_p \frac{n-1}{n} \frac{[SI][IT]}{[I]} - [SI]- c_t \frac{(n-1)(n-2)}{n^2} \frac{[SI][IT]^2}{[I]^2}\\
{[\dot{S T}]} & = -\tau \frac{n-1}{n} \frac{[SI][ST]}{[S]} + q [SI] + c_p \frac{n-1}{n} \frac{[SI][IT]}{[I]} - a [ST] + c_t \frac{(n-1)(n-2)}{n^2} \frac{[SI][IT]^2}{[I]^2}\\
{[\dot{I I}] } & = 2 \tau ([S I] + \frac{n-1}{n} \frac{[SI]^2}{[S]}) -2 [I I] - 2 c_p \frac{n-1}{n} \frac{[II][IT]}{[I]} - 2 c_t \frac{(n-1)(n-2)}{n^2} \frac{[II][IT]^2}{[I]^2}\\
{[\dot{I T}] } & = \tau \frac{n-1}{n} \frac{[SI][ST]}{[S]} - a [I T] + c_p(\frac{n-1}{n} \frac{[II][IT]}{[I]} - \frac{n-1}{n} \frac{[IT]^2}{[I]} - [I T]) + q [II] - [IT] + \\ & \quad \quad \quad + c_t \frac{(n-1)(n-2)}{n^2} \frac{[II][IT]^2}{[I]^2} - c_t (\frac{n-1}{n} \frac{[IT]^2}{[I]} + \frac{(n-1)(n-2)}{n^2} \frac{[IT]^3}{[I]^2}),\\
{[\dot{T T}]} & = 2 q [I T] + 2 c_p ([IT] + \frac{n-1}{n} \frac{[IT]^2}{[I]}) + 2 c_t (\frac{n-1}{n} \frac{[IT]^2}{[I]} + \frac{(n-1)(n-2)}{n^2} \frac{[IT]^3}{[I]^2}) - 2 a [T T],
\end{cases}
\end{align}
where the transition rates have been renormalized by the natural timescale $g + h$: $$\tau \leftarrow \frac{\tau}{g+h},\ c_p \leftarrow \frac{c_p}{g+h},\ c_t \leftarrow \frac{c_t}{g+h},\ a \leftarrow \frac{a}{g+h},\mbox{ and }q:= \frac{g}{g+h}. $$ From here onward, we only consider the time-scaled transition rates.

When $q = 0$, natural recovery dominates over the transition to the treated state, yielding the classical SIR model. Similarly, when $a \rightarrow \infty$, individuals reside in the treated state for a vanishing duration, and we recover the SIR model. Instead, when $q \rightarrow 1$ (no bypass of the treated state $T$) and $c_t = 0$ (no triplewise contact tracing), we recover the original model by Eames and Keeling~\cite{eames2003contact}.



The usual way of calculating threshold values of parameters is to evaluate the leading eigenvalue of the Jacobian of the system at the disease-free steady-state. For the pairwise SITR model [Eq.~\eqref{eq:system}], this approach is not directly applicable, since the closure relations become ill-defined at the disease-free steady-state due to the presence of $[I] = 0$ in the denominator. Nevertheless, the threshold can be identified numerically, as demonstrated in Appendix~\ref{sec:appendixB}. Moreover, in what follows, we present a mathematical method based on fast variables that is able to capture the threshold behaviour analytically, yielding threshold values that agree well with the numerical results.

\section{Results}
In this section, we show that although incorporating a local network structure renders standard linear stability analysis unable to determine the critical contact tracing threshold, this limitation can be overcome by exploiting correlations emerging at early times.

Indeed, we first show that early-time correlations govern a reformulation of the epidemic threshold problem in terms of so-called `fast variables'. 
This alternative approach reveals that the contact tracing benefits from a network structure when it comes to its contribution to outbreak eradication.
We then show analytically how the success of contact tracing heavily depends on the density 
of the network structure, the contact tracing coverage level, and its timescale relative to disease progression.

\subsection{Epidemic threshold via fast variables}

In linear stability analysis, the epidemic threshold is defined by the critical parameter values at which the disease-free steady state loses stability. Equivalently, this threshold corresponds to the point in parameter space where the growth rate of the infected population vanishes. The condition of zero growth therefore provides a necessary and sufficient criterion for the epidemic threshold. In the equation for the growth of the number of infected individuals $[\dot{I}]$ in Eq.~\eqref{eq:system}, 
$[I]$ can be factorized to give:
\begin{equation}
    \frac{[\dot{I}] }{[I]} = \tau\frac{[S I]}{[I]} - 1 - c_p \frac{[I T]}{[I]} - c_t \frac{n-1}{n} \left(\frac{[IT]}{[I]}\right)^2,
\end{equation}
and the vanishing of the right hand side separates regimes of exponential growth and decay of the number of infected individuals. 
Based on this reformulation, the epidemic threshold depends on the limiting values of $[SI]/[I]$ and $[IT]/[I]$.

Motivated by this observation, we define the new variables $x(t):= [SI]/[I]$ and $z(t):=[IT]/[I]$. To find their limiting values, we set down the equations for their time evolution $\dot x = \dot{[SI]}/[I] - x \dot{[I]}/[I]$ and $\dot z = \dot{[IT]}/[I] - z \dot{[I]}/[I]$. Writing the equations for $\dot{[SI]}/[I]$, $\dot{[IT]}/[I]$, and $\dot{[I]}/[I]$ using Eq.~\eqref{eq:system} eventually requires to define and find the limiting values of the following set of new variables (see Appendix~\ref{sec:appendixA}):
\begin{equation}
\label{eq:def_fast}
    x(t)=\frac{[SI]}{[I]}, \quad z(t)=\frac{[IT]}{[I]}, \quad \text{and} \quad y(t):=\frac{[II]}{[I]}, \quad  u(t):=\frac{[SS]}{[S]} , \quad  v(t):=\frac{[SI]}{[S]}  , \quad  w(t):=\frac{[ST]}{[S]}.
\end{equation}

\begin{figure}
    \centering
    \includegraphics[width=0.8 \textwidth]{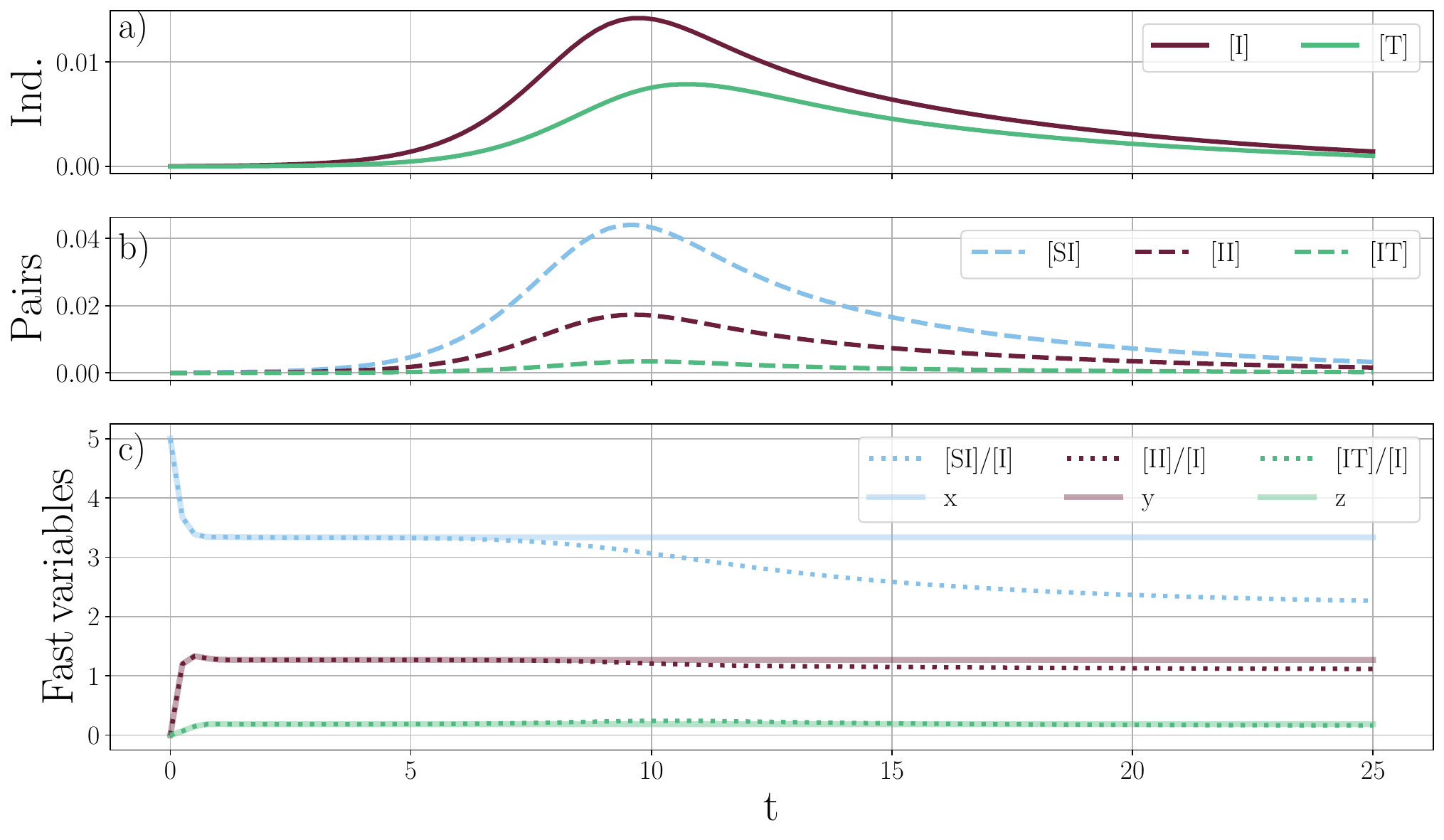}
    \caption{
    \textbf{The fast variables reach a transient equilibrium much sooner compared to the time scale of epidemic evolution}. a) Time evolution of $[I]$ and $[T]$. b) Time evolution of the pairs $[SI]$, $[II]$, and $[IT]$. c) Time evolution of the fast variables $x$, $y$, and $z$, computed from the full system in Eq.~\eqref{eq:system} (dotted line) or from the fast-variables system in Eq.~\eqref{eq:fast_var_sys} (solid line). Parameter values: $n = 5$,
$\tau = 1.1$, $a = 1$, $q = 0.6$, $c_p = 10$, and $c_t = 0$ (pure pairwise). Initial conditions: $[I](t=0)=[T](t=0) = 10^{-5}$.}
    \label{fig:fast_variables}
\end{figure}

Barnard et al. show in~\cite{barnard2019epidemic} that, as illustrated in Fig.~\ref{fig:fast_variables}, these so-called `fast' variables reach a transient equilibrium much faster compared to the time scale of epidemic evolution. Indeed, early on in an epidemic a correlation structure is formed that drives this early saturation of variables. `Waiting' for the epidemic to `feel' the network turns out to be the only reasonable way to measure the epidemic threshold.

Overall, the epidemic threshold can be obtained by identifying the conditions on the model parameters that cancel the growth of the prevalence $\dot{[I]}$ once the fast variables have reached their steady state. We thus need to solve:
\begin{equation}
\label{eq:threshold_system1}
    \dot x = 0, \quad \dot z = 0, \quad \dot y = 0, \quad \dot u = 0, \quad \dot v = 0, \quad \dot w = 0, \quad \text{and} \quad \frac{\dot{[I]}}{[I]} = 0.
\end{equation}

\paragraph{Proximity with the initial condition.}

Assuming proximity with the initial condition, we can impose $u$ constant and equal to the average degree $n$, which holds as long as the initial epidemic seed is small and mixes randomly with the susceptible population: 
$[SS] (0) \sim n N$ and $[S](0) \sim N$. Initially, we can also assume that both $v$ and $w$ are constant and equal to zero, since $[SI] (0) \sim [ST] (0) \sim 0$.

This approximation reduces Eq.~\eqref{eq:threshold_system1} to:
\begin{equation}
\label{eq:fast_var_sys_+}
    \dot x = 0, \quad \dot y = 0, \quad \dot z = 0, \quad \frac{\dot{[I]}}{[I]} = 0,
\end{equation}
where, using Eq.~\eqref{eq:system}, we have
\begin{align}
\label{eq:fast_var_sys}
\begin{cases}
\dot x &= \frac{\dot{[SI]}}{[I]} - x \frac{\dot{[I]}}{[I]}\\
&= \tau x (n-2) + \frac{c_p}{n} x z + 2 \frac{c_t}{n} \frac{n-1}{n} x z^2 - \tau x^2,\\
\dot y &= \frac{\dot{[II]}}{[I]} - y \frac{\dot{[I]}}{[I]}\\
&= \tau x (2 - y) - y - \frac{n-2}{n} c_p y z - \frac{n-1}{n} \frac{n-4}{n} c_t y z^2,\\
\dot z &= \frac{\dot{[IT]}}{[I]} - z \frac{\dot{[I]}}{[I]}\\
&= - \tau xz - (a+ c_p) z + c_p \frac{n-1}{n} y z + \frac{c_p}{n} z^2 + q y + c_t \frac{(n-1)(n-2)}{n^2} y z^2 - c_t \frac{n-1}{n} z^2 + c_t \frac{2}{n}\frac{n-1}{n} z^3.
\end{cases}
\end{align}
Eventually, the epidemic threshold can be computed from Eq.~\eqref{eq:fast_var_sys_+} or equivalently from:
\begin{equation}
\label{eq:tobesolved}
\frac{[\dot{I}]}{[I]} = 0, \quad \frac{[\dot{SI}]}{[I]} = 0, \quad \frac{[\dot{II}]}{[I]} = 0, \quad \frac{[\dot{IT}]}{[I]} = 0.
\end{equation}

Using the definition of the fast variables [Eq.~\eqref{eq:def_fast}] and dividing the equations for $\dot{[SI]}$, $\dot{[II]}$, and $\dot{[IT]}$ in Eq.~\eqref{eq:system} by $[I]$, Eq.~\eqref{eq:tobesolved} finally reads:
\begin{align}
\label{eq:tobesolved_explicit}
\begin{cases}
& \tau x - 1 - c_p z - c_t \frac{(n-1)}{n} z^2 = 0\\
& \tau x (n-2)-c_p\frac{(n-1)}{n} x z - x -c_t \frac{(n-1)(n-2)}{n^2} x z^2 = 0\\
& 2 \tau x -2 y -2 c_p \frac{(n-1)}{n} y z - 2 c_t \frac{(n-1)(n-2)}{n^2} y z^2 = 0\\
& -a z + c_p \left(\frac{(n-1)}{n} y z- \frac{(n-1)}{n} z^2-z\right) + (q y - z) + \\ & \quad \quad \quad +c_t \frac{(n-1)(n-2)}{n^2} y z^2 - c_t \frac{(n-1)}{n} z^2 - c_t \frac{(n-1)(n-2)}{n^2} z^3 = 0.
\end{cases}
\end{align}
Here, contact tracing, through its parameters $c_p$ and $c_t$, influences the derivation of the epidemic threshold, in contrast to the homogeneous mean-field formulation of the model, where it has no effect (see Appendix~\ref{sec:homogeneous}). 
We can thus compute the level of contact tracing ($c_p$ or $c_t$) that is required to bring the system to the disease-free steady-state. 
In Appendix~\ref{sec:appendixB}, we show through numeric solving of the ODE system in Eq.~\eqref{eq:system} that the critical level of contact tracing computed in this way separates a regime where the number of infected individuals in the population grows or decays exponentially fast, once the fast variables have reached a transient steady-state. There, we also show that the critical level of tracing also separates regimes where a zero or nonzero final epidemic size (stationary number of recovered individuals) can be reached with an infinitesimal infected seed.

\subsection{Critical thresholds for control}

Leveraging early time correlations with the help of fast variables, we were able to reformulate the epidemic threshold problem in terms of growth and decay of the population of infected individuals. 
We found that, when accounting for a local network structure through a pairwise representation, the contact tracing parameters affect the capacity of the system to reach the disease-free steady-state and thus, we can look for a critical level of contact tracing. 

In this section, we first analyze the pure pairwise and pure triplewise scenarios separately. We then discuss the differences between the two distinct implementations of contact tracing and the consequences of combining them. 

\subsubsection{Pairwise contact tracing}

First, we consider that an individual in state $T$ can independently influence each of its infected neighbors in $I$ into seeking treatment. This mechanism of pairwise social influence corresponds for example to the digital and manual implementations of forward contact tracing in an airborne disease outbreak, whereby confirmed cases who wear face masks inform their contacts about their infected status, eventually influencing them into considering adoption of mitigation measures too.


The critical level of contact tracing can be derived from Eq.~\eqref{eq:tobesolved_explicit} (more details in Appendix~\ref{sec:appendix_derivation}), by identifying the pure pairwise contact tracing parameter $c_p$ that solves the system, under the constraint of no triplewise contact tracing, i.e. $c_t = 0$.
In this case, the critical level of pairwise contact tracing required to eradicate the spreading disease is:
\begin{equation}
\label{eq:threshold_purepairwise}
    c_p^* = \left(a + R_0\right) \frac{R_0 - 1}{F_p(n, q, \tau)},
\end{equation}
where $F_p(n, q, \tau):= \frac{1}{n} \frac{R_0(nq - 1)+ (1-q)}{R_0}$, and $R_0 := \tau (n-2)$ is the basic reproductive number (note that $\tau$ is the ratio between the infection rate and the total recovery rate). 

\begin{figure}
    \centering
    \includegraphics[width=16cm]{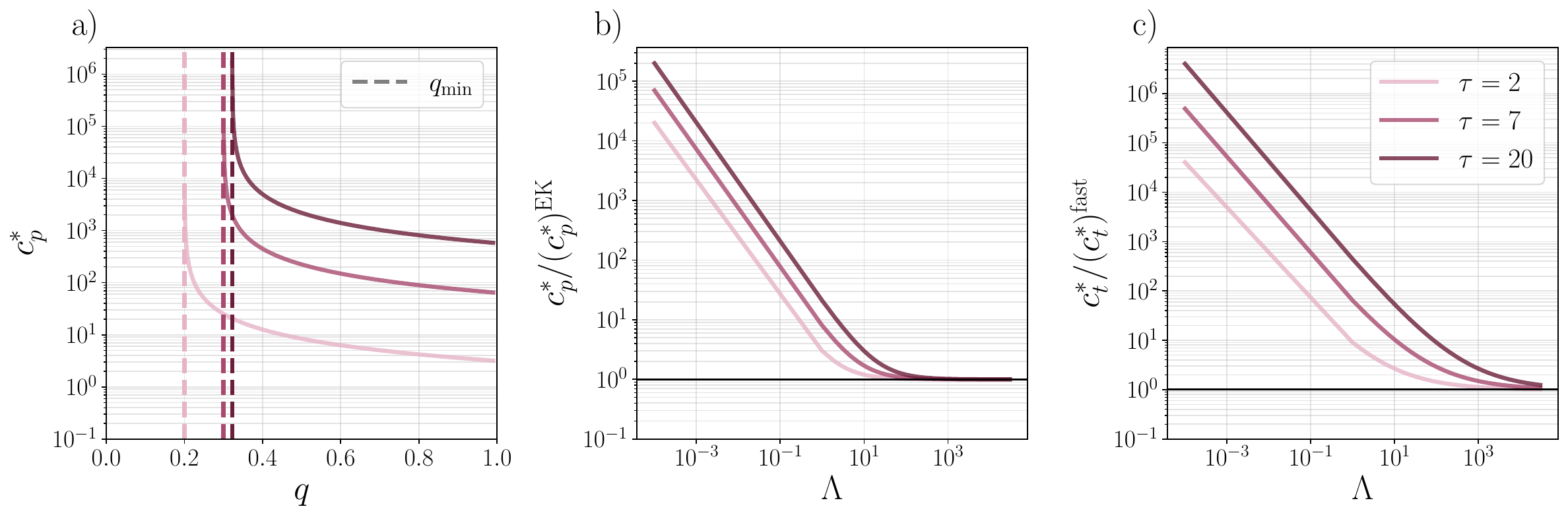}
    \caption{
    \textbf{The threshold diverges as a function of tracing coverage and slowness of CT.} a) Critical level of pairwise CT as a function of the tracing coverage $q$, for three values of the transmission rate: $\tau = 2, 7, 20$. The dashed vertical lines represent $q_{\text{min}}^p$. Parameter value: $a = 0.1$. b) Ratio between the pure pairwise critical level of CT computed with the method of fast variables ($c_p^*$) and the Eames and Keeling threshold (EK), as a function of the celerity of CT, through $\Lambda$. We consider the same three values of transmission rate as in panel a, and $c_t = 0$. c) Ratio between the pure triplewise critical level of CT computed with the method of fast variables ($c_t^*$) and the pure triplewise critical level obtained in the adjusted fast tracing limit, as a function of the celerity of CT, through $\Lambda$. We consider the same three values of transmission rate as in panel a and $c_p = 0$. Parameter values for panels b and c: $q = 1$, and $a = 1$. Parameter values common across the three panels: $n=3$. 
    }
    \label{fig:threshold}
\end{figure}

When the compliance gets large $q\rightarrow 1$, $F_p$ satisfies $F_p(n, 1, \tau) = \frac{n - 1}{n}$, which is the pre-factor of the closure of triples. 
Instead, a lower bound $q_{\text{min}}^p \leq q$ on the compliance with treatment or tracing coverage is imposed by the condition $c_p^* \geq 0$, with $q_{\text{min}}^p = \frac{R_0 - 1}{R_0 n - 1}$.
When the coverage approaches its lower bound $q \rightarrow (q_{\text{min}}^p)^+$, $F_p \rightarrow 0$ and the critical level of contact tracing $c_p^* \rightarrow \infty$ diverges. Thus, when tracing coverage becomes too small $q < q_{\text{min}}$, no finite level of pairwise CT is able to contain an epidemic. In Fig.~\ref{fig:threshold}a, we see for several values of the spreading rate $\tau$,  how the critical level of pairwise tracing diverges when $q \rightarrow (q_{\text{min}}^p)^+$ and then decays with $q > q_{\text{min}}^p$. Although the tracing coverage is expected to affect the performance of contact tracing, we here show that a minimum coverage needs to be reached for contact tracing to be able to contribute to suppressing an outbreak. Moreover, the impact of the tracing coverage is such that a change of $20\%$ in its magnitude can require a $10$ times higher level of contact tracing.

We stress that the critical level of contact tracing in Eq.~\eqref{eq:threshold_purepairwise} is obtained without working under the unrealistic fast tracing assumption used in the seminal work of Eames and Keeling on contact tracing~\cite{eames2003contact}.
For a direct comparison with the result of Eames and Keeling in~\cite{eames2003contact}, we recall that their model considers $q = 1$, so also $F_p = \frac{n-1}{n}$. 
In addition, fast contact tracing is imposed by scaling the contact rates $c_p$ and $a$ as $\Lambda c_p$ and $\Lambda a$, with $\Lambda \gg 1$, while leaving all other rates (disease progression rates) unchanged. In this unrealistic limit then, the threshold reduces to:
\begin{align}
\label{eq:EK_threshold}
    c_p^* &= a\left(1 + \frac{R_0}{\Lambda a}\right) \frac{R_0 n - n}{n-1} \nonumber\\
    &= a \frac{n (R_0 - 1)}{n-1} = (c_p^*)^{\text{EK}},
\end{align}
which is the threshold of Eames and Keeling and disregards the second term in Eq.~\eqref{eq:threshold_purepairwise}. 

In Fig.~\ref{fig:threshold}b, we show how the fast variables threshold of Eq.~\eqref{eq:threshold_purepairwise} compares with the Eames and Keeling threshold of Eq.~\eqref{eq:EK_threshold}, when $q = 1$ and for different values of $\Lambda$. 
When CT is slow compared to the disease progression and $\Lambda \ll 1$, we see that the actual level of tracing required to eradicate a disease can be orders of magnitude higher than what expected under the fast tracing assumption. As expected, the agreement improves when $\Lambda$ grows.
In the fast tracing limit, the critical level of tracing depends linearly on the basic reproductive number and is affected by the density (through $n$) and the period of active tracing (through $a$). When contact tracing becomes progressively slower, the quadratic dependence on the basic reproductive number becomes important, eventually taking over, and the dependence on $a$ disappears. 

Overall, Fig.~\ref{fig:threshold} shows that low tracing coverage and slow contact tracing compared to the timescale of disease progression can significantly undermine the efficacy of the contact tracing measure in bringing the system to the disease-free steady-state. 

\begin{figure}
    \centering
    \includegraphics[width=16cm]{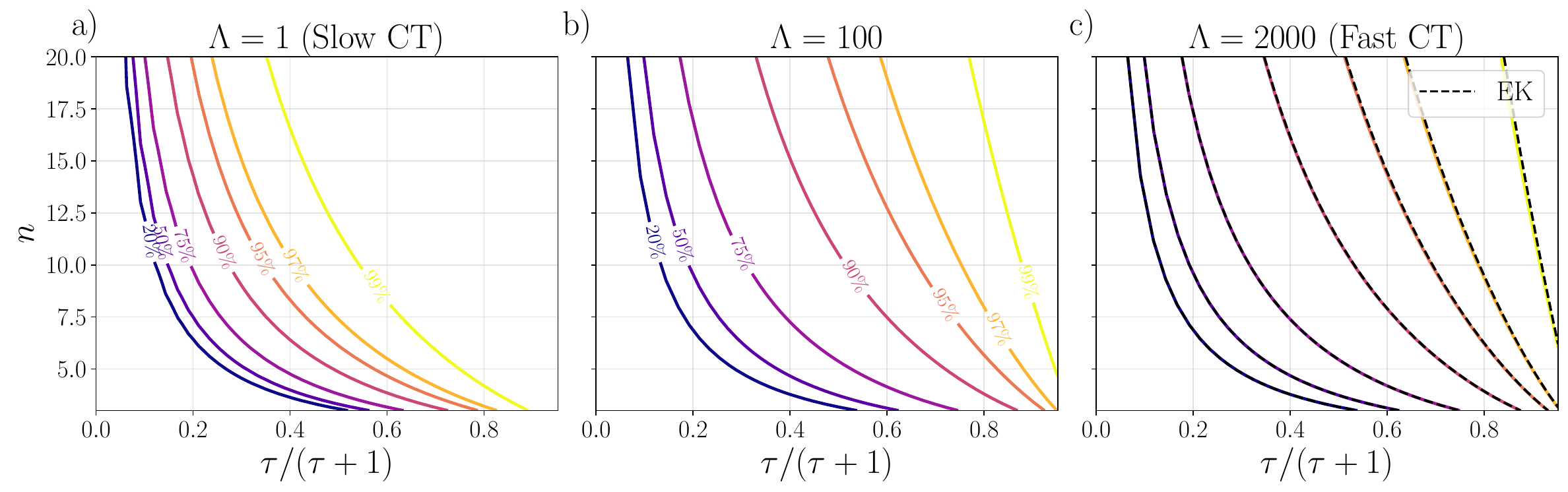}
    \caption{\textbf{The social weight of CT increases dramatically for dense networks, as we move away from the fast tracing limit.} Contour lines of the (analytical) social weight $(c_p^*/a)/((c_p^*/a) + 1)$ in the following plane: network density $n$ vs probability that infection passes across a contact $\tau/(\tau + 1)$. a) Fast variables result when $\Lambda = 1$. b) Fast variables result when $\Lambda =  100$. c) Eames and Keeling's (EK) fast tracing result (dashed black lines) superimposed on the fast variables result when $\Lambda = 2000$. Parameter values: $c_t = 0$, $a=1$, and $q=1$.}
    \label{fig:contour}
\end{figure}

In Fig.~\ref{fig:contour}, we show that also the network density significantly affects the containment capacity of contact tracing, especially away from the fast contact tracing limit.
Similarly to~\cite{eames2003contact}, we let $\tau/(\tau + 1)$ be the probability that infection passes across a contact and $(c_p^*/a)/((c_p^*/a) + 1)$ the proportion of contacts that need to be traced, i.e., the social weight of the CT measure. 
In the fast tracing limit, the social weight is independent of the period of tracing activity (through $a$).
However, stepping out of the fast tracing limit by decreasing $\Lambda$ reveals a dependence of the social weight on $a$, that reshapes the social weight landscape in the space of parameters. Crucially, the dependence of the social weight on the network density is amplified when contact tracing gets slower. 

\subsubsection{Triplewise contact tracing}
Formal implementations of contact tracing through digital applications or healthcare manual efforts often require a single pairwise interaction to trigger a contact tracing notification and to encourage the traced individual to seek treatment. 
In this sense, formal implementations of contact tracing are relatively well captured by the pairwise implementation described in the previous section~\cite{barrat2021effect}.

Nevertheless, beyond such formal implementations, contact tracing can also arise informally when awareness of an outbreak spreads within a population. In informal settings, individuals influence one another to exchange information about potential exposures and to adopt protective behaviors, generating tracing-like effects through social interactions. This spontaneous, decentralized contact tracing closely resembles opinion-dynamics–driven behavioral adoption \cite{yasseri_opinion_2025,starnini_opinion_2025} and may require social reinforcement through exposure to multiple contacts. In this section, we study the pure triplewise implementation of contact tracing, where individuals only get traced if they have been in contact with at least two contact tracing-triggering nodes. This implementation of contact tracing may also describe the adoption of contact tracing by individuals who are particularly reluctant to engage with the process.

Solving Eq.~\eqref{eq:tobesolved_explicit} (details in Appendix~\ref{sec:appendix_derivation_triplewise}) for $c_t$ with $c_p = 0$ (no pairwise contact tracing), we find that the critical level of triplewise contact tracing required to eradicate the spreading infection is: 

\begin{equation}
\label{eq:threshold_triplewise}
    c_t^* = \left(a + R_0\right)^2 \frac{R_0 - 1}{F_t(n, q, \tau)},
\end{equation}
where $F_t(n, q, \tau) = \frac{(n-1)}{n^2(n-2)} \left(\frac{R_0 (n q - 2) + 2(1-q)}{R_0}\right)^2$. 

When the tracing coverage becomes large $q\rightarrow1$, $F_t$ satisfies $F_t(n, 1, \tau) = \frac{(n-1)(n-2)}{n^2}$, which is the pre-factor of the closure of 3-stars. 
Moreover, we have a lower bound on the compliance with treatment: $q \geq q_{\text{min}}^t = 2 \frac{R_0 - 1}{R_0 n - 2}$,
and when the tracing coverage approaches its lower bound $q \rightarrow (q_{\text{min}}^t)^+$, $F_t \rightarrow 0$ and the critical level of CT $c_t^* \rightarrow \infty$ diverges.

The critical level of triplewise tracing exhibits a similar structure to the one of pairwise tracing, with a dependence on tracing coverage, network density, and speed of contact tracing.
However, in this case, if we define the fast tracing limit in an analogous way to Eames and Keeling (scaling both the contact rates $c_p$ and $a$ as $\Lambda c_p$ and $\Lambda a$, with $\Lambda \gg 1$) would make the critical level of triplewise contact tracing diverge. Indeed, when contact tracing is too fast, contact-tracing-triggering nodes terminate their tracing activity so fast that $[TIT]$ triples become too rare and contact tracing fails.
However, we can define an alternative and more meaningful fast tracing limit for this form of contact tracing by scaling $a$ as $\sqrt{\Lambda} a$, where $\Lambda \gg 1$ is the scaling of $c_t \rightarrow c_t \Lambda$. This fast tracing limit adjusted to the features of triplewise contact tracing is characterized by an enhanced longevity of active tracing. In Fig.~\ref{fig:threshold}c, we show that in this adjusted fast tracing limit, as long as CT is sufficiently fast, it is possible to neglect the $R_0$ next to $a$ in Eq.\eqref{eq:threshold_triplewise}, similarly to what we observed for pure pairwise CT (panel b).

\subsubsection{Epidemic control under combined tracing interventions}

Naturally, during an outbreak, the formal and informal implementations of contact tracing can coexist.
In this section, we compare the efficacy of the pure pairwise and triplewise implementations of contact tracing before investigating the consequence of combining them.

\begin{figure}
    \centering    \includegraphics[width=16cm]{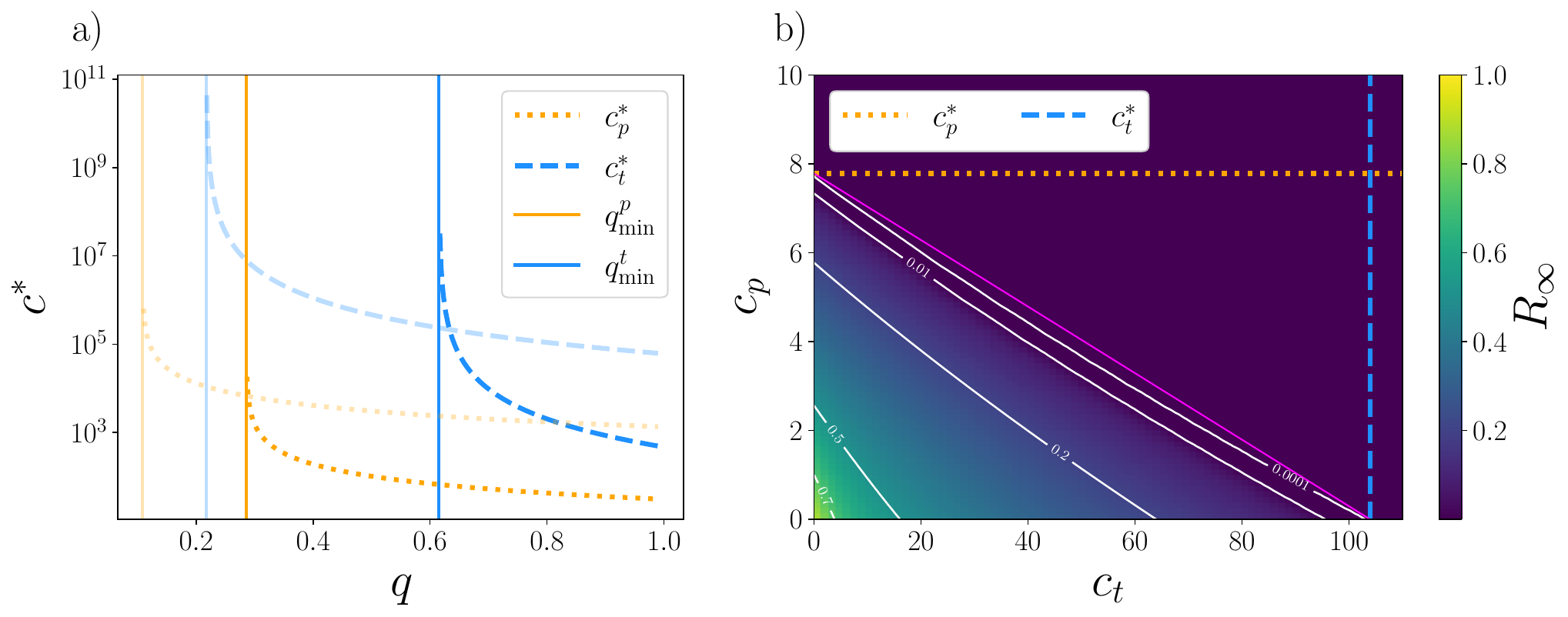}
    \caption{\textbf{Triplewise CT is more stringent and less efficient than pairwise CT.} a) Analytical critical level of CT as a function of the tracing coverage $q$ for both pairwise (dotted, orange) and triplewise (dashed, blue) contact tracing. The vertical solid lines indicate the minimum compliance $q_{\text{min}}$. Parameter values: $\tau = 5$, $a = 0.1$, and two choices of the network density, $n=3$ (full opacity) and $n=9$ (partial transparency). b) Heatmap and contour lines of the final epidemic size $R_{\infty}$ in the following plane: $c_p$ vs $c_t$. The dotted orange and dashed blue lines are the analytical critical levels of contact tracing for pure pairwise and pure triplewise CT, respectively. Parameter values: $n = 5$, $\tau = 0.7$, $a = 1$, $q = 0.6$. Initial conditions: $[I](0) = [T](0) =5 \cdot 10^{-8}$.}
\label{fig:comparison}
\end{figure}

In Fig.~\ref{fig:comparison}a we show, for a representative choice of the parameters, that pure triplewise contact tracing always requires higher rates of contact tracing than pure pairwise, to bring the system to the disease free equilibrium. Additionally, pure triplewise CT systematically needs higher tracing coverages in order to be efficient. Although the critical level of CT grows for denser networks, network density helps expand the range of coverages that are sufficient for the contact tracing measure to be efficient.

Finally, the combination of both types of contact tracing only results in a slight reinforcement of contact tracing, which can be attributed to the correlations between pairwise and triplewise contact tracing. 
In Fig.~\ref{fig:comparison}b, we show how the final epidemic size varies for different combinations of the contact tracing rates $c_p$ and $c_t$. As mentioned above, the critical rates of contact tracing coincide with the rates that separate a vanishing from a non-vanishing final epidemic size.
If both types of contact tracing were independent, we would expect the critical levels of contact tracing $c_p$ and $c_t$ to lie on the line connecting the threshold of pure pairwise and pure triplewise contact tracing (solid, magenta line). Instead, we find that when the contact tracing rates of each type -$c_p$ and $c_t$- are intermediate, the final epidemic size is smaller than would be expected. 

\section{Conclusion}
\label{sec:results}

In a seminal study by Eames and Keeling for pairwise contact tracing on networks~\cite{eames2003contact}, an analytical expression was derived for the critical level of contact tracing required to suppress an outbreak, under the strong assumption that contact tracing occurs on a timescale much faster than disease progression. This assumption may be reasonable for sexually transmitted infections (STI), for which most major contact tracing programs until the COVID-19 pandemic were deployed~\cite{brandt2022history, hossain2022effectiveness}. Indeed, STIs tend to be associated with long infectious periods~\cite{stephenson2023sharing}, compared to typical times of deployment of contact tracing efforts. However, the assumption of fast contact tracing fails for infectious diseases in general and for airborne infections like COVID-19, in particular. 

In this paper, we crucially showed that, by exploiting correlations that arise at early times through the method of fast variables introduced in~\cite{barnard2019epidemic}, we could overcome both the limitations of standard linear stability analysis in pairwise models and the difficulty of determining critical tracing levels in the general case, when the fast-tracing assumption no longer holds.
To do so, we analyzed the ability of contact tracing to drive an epidemic system toward the disease-free state in a pairwise framework. 
The pairwise approximation allows us to consider some level of (local) network structure, which is crucial to study an intervention that leverages contact patterns, while keeping the model analytically tractable. 
Within this setting, we examined two implementations of the intervention: one based on pairwise interactions and another involving triplewise interactions with contact-tracing–triggering nodes. These correspond, respectively, to formal and informal implementations, or to less and more reluctant adoptions of forward contact tracing. To account for evidence on the impact of contact tracing app coverage on the efficacy of the measure, we also considered that only part of the infected populations has the ability to become contact-tracing triggering nodes~\cite{barrat2021effect, moreno2021anatomy, mancastroppa2021stochastic}. 

Additionally, our results show that network density non-linearly influences the effectiveness of contact tracing. 
While both pairwise and triplewise contact tracing become less efficient as networks become denser (they require a higher critical rate of contact tracing), sufficiently sparse networks are also unfavorable, as they increase the minimum tracing coverage required for contact tracing to be effective, thereby narrowing the range of coverages over which it can contribute.
Moreover, by deriving results in full generality, rather than restricting attention to the limit in which contact tracing is much faster than disease progression, we could demonstrate that assuming fast contact tracing can substantially underestimate the required tracing effort and overestimate the efficacy of the intervention. The intervention should be deployed at a pace that is at least comparable to the spreading rate of the disease.
Finally, although triplewise contact tracing —representing a more spontaneous or informal implementation of the measure— is always less efficient than its pairwise counterpart, it can still make a meaningful contribution, especially when the spread of protective behaviors through contact tracing occurs on a timescale comparable to that of disease progression or as long as the tracing activity of treated individuals lasts sufficiently long.

To maintain analytical tractability, we deliberately restricted the number of epidemic states, omitting compartments that are often important for more realistic representations of disease progression, such as an exposed state for influenza-like illnesses~\cite{chowell2008seasonal, trentini2022characterizing}. The assumption that contact tracing triggering nodes are no longer infectious is also a strong simplification. Future work should investigate the effect of allowing contact tracing triggering nodes to continue contributing to pathogen spread. Indeed, in practice, health authorities typically implement contact tracing by identifying and testing the contacts of confirmed cases, with mitigation measures adopted only by those whose diagnostic test results return positive. 
Moreover, despite incorporating some level of network representation, we also neglected numerous features of real contact networks such as clustering~\cite{house2010impact}, higher-order interactions, distinct network layers for contact tracing and disease progression~\cite{mancastroppa2021stochastic}, and temporal variations in network links~\cite{holme_temporal_2012,newman_networks_2018,kivela_multilayer_2014,unicomb_dynamics_2021}.
Finally, while our analysis focuses on how contact tracing shifts the epidemic threshold, this intervention can also shape the epidemic curve by affecting peak height and timing. These effects are crucial for determining whether the health-care system becomes saturated and for buying time to deploy resources such as diagnostic tests, face masks, and vaccination campaigns to further mitigate the outbreak~\cite{di2020impact}.

\subsection*{Competing interests}
We declare that we have no competing interests.

\subsection*{Acknowledgments}
 P.L. Simon acknowledges support from the National Research, Development and Innovation Office in Hungary (RRF-2.3.1-21-2022-00006).

\newpage
\bibliographystyle{unsrt}
\bibliography{project1.bib}
\newpage

\appendix

\section{Full system of fast variables}
\label{sec:appendixA}

In general, the evolution of $x$ and $z$ is governed by:
\begin{align}
\begin{cases}
\dot x &= \frac{\dot{[SI]}}{[I]} - x \frac{\dot{[I]}}{[I]}\\
&= \tau x (\frac{n-1}{n} (u-v)-1) + \frac{c_p}{n} x z + 2 \frac{c_t}{n} \frac{n-1}{n} x z^2 - \tau x^2,\\
\dot z &= \frac{\dot{[IT]}}{[I]} - z \frac{\dot{[I]}}{[I]}\\
&= \tau x (\frac{n-1}{n} w - z) - (a + c_p) z + c_p \frac{n-1}{n} y z + \frac{c_p}{n} z^2 + q y + c_t \frac{(n-1)(n-2)}{n^2} y z^2 - c_t \frac{n-1}{n} z^2 + c_t \frac{2}{n}\frac{n-1}{n} z^3,\\
\end{cases}
\end{align}
where $u:= \frac{[SS]}{[S]}$, $v:=\frac{[SI]}{[S]}$, $w:= \frac{[ST]}{[S]}$, and $y:= \frac{[II]}{[I]}$.

We thus also need to write down the evolution of these variables, which reads:
\begin{align}
\label{eq:full_fast_var_sys}
\begin{cases}
\dot u &= \frac{\dot{[SS]}}{[S]} - u \frac{\dot{[S]}}{[S]}\\
&= -\tau(2 \frac{n-1}{n} + 1) u v,\\
\dot v &= \frac{\dot{[SI]}}{[S]} - v \frac{\dot{[S]}}{[S]}\\
&= \tau \frac{n-1}{n} u v + \frac{\tau}{n} v^2 - (\tau - 1)v - c_p \frac{n-1}{n} v z - c_t  \frac{(n-1)(n-2)}{n^2} v z^2,\\
\dot w &= \frac{\dot{[ST]}}{[S]} - w \frac{\dot{[S]}}{[S]}\\
&= \frac{\tau}{n} v w + q v - aw + c_p \frac{n-1}{n} vz + c_t \frac{(n-1)(n-2)}{n^2} v z^2,\\
\dot y &= \frac{\dot{[II]}}{[I]} - y \frac{\dot{[I]}}{[I]}\\
&= \tau x (2(1+\frac{n-1}{n} v) - y) - y - \frac{n-2}{n} c_p y z - \frac{n-1}{n} \frac{n-4}{n} c_t y z^2.
\end{cases}
\end{align}

Assuming proximity with the initial condition implies considering $u$, $v$, and $w$ as constant and more specifically $u=n$ and $v = w = 0$. This assumption yields the following system of equations for the evolution of fast variables:
\begin{align}
\begin{cases}
\dot x = \tau x (\frac{n-1}{n} (n-0)-1) + \frac{c_p}{n} x z + 2 \frac{c_t}{n} \frac{n-1}{n} x z^2\\
\dot z = \tau x (0 - z) - (a+ c_p) z + c_p \frac{n-1}{n} y z + \frac{c_p}{n} z^2 - q z + c_t \frac{(n-1)(n-2)}{n^2} y z^2 - c_t \frac{n-1}{n} z^2 + c_t \frac{2}{n}\frac{n-1}{n} z^3\\
\dot u = 0\\
\dot v = 0\\
\dot w = 0\\
\dot y = \tau x (2(1+0) - y) - y - \frac{n-2}{n} c_p y z - \frac{n-1}{n} \frac{n-4}{n} c_t y z^2.
\end{cases}
\end{align}

An SIS version of this model would only require the addition of a $+a z$ term in the $\dot{[SI]}/[I]$ equation.

\section{Homogeneous mean-field approximation}
\label{sec:homogeneous}

Under the homogeneous mean-field approximation, we can write the ordinary differential equations (ODE) describing the time evolution of the number of nodes in each state:

\begin{align}
\label{eq:meanfield}
\begin{cases}
{\dot{S} } & =-\tau I S \\
{\dot{I}} & = +\tau I S - (g + h)I - c_p T I  - c_t I T^2\\
{\dot{T} } & = g I + c_p T I - a T + c_t I T^2 \\
{\dot{R} } & = a T + h I.
\end{cases}
\end{align}

Identifying the epidemic threshold in this context corresponds to finding the condition on the model parameters that makes the leading eigenvalue of the Jacobian at the disease-free steady-state equal to zero. This condition separates a regime where the disease-free steady-state is locally stable from one where it is not.
Linearizing Eq.~\eqref{eq:meanfield} around the disease-free state immediately eliminates all contributions of contact tracing (through its parameters $c_p$ and $c_t$) to the epidemic threshold, thereby suggesting that contact tracing alone is unable to eradicate a disease.

The literature proposes a variety of modelling choices to represent the effects of contact tracing, and these choices can influence whether the intervention contributes to the epidemic threshold or not. For instance, in~\cite{demeijere2023}, traced individuals are defined as those infected by a contact tracing–triggering node. When, as per this modelling choice, contact tracing depends on the pairwise interaction between susceptible and infected individuals, the linear stability analysis no longer cancels the contribution of contact tracing to the epidemic threshold.

\section{Numerical threshold}
\label{sec:appendixB}

The epidemic threshold can be studied numerically in two different ways: through the final epidemic size or the initial growth rate of the epidemic.

\paragraph{Final epidemic size.}

\begin{figure}
    \centering
    \includegraphics[width=16cm]{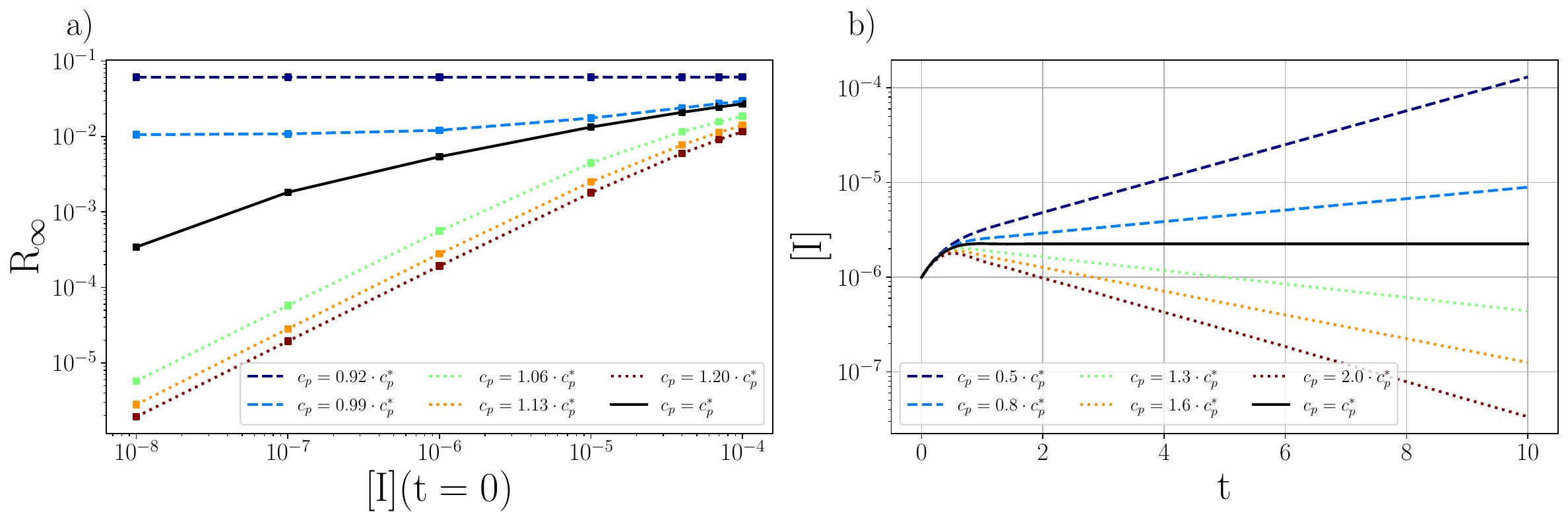}
    \caption{\textbf{Numerical threshold in the pure pairwise contact tracing.} a) The final epidemic size $R(\infty)$ as a function of the initial condition $I(0)$ for different values of the tracing parameter $c_p$. b) The time dependence of the prevalence $[I]$ for different values of the contact tracing rate $c_p$ with initial condition $I(0)=10^{-6}$. 
    Parameter values: $n = 5$, $a = 1$, $p = 0.6$, $c_p = 0.3$, and $c_t = 0$.}
    \label{fig:finalsize_I0}
\end{figure}

In terms of the final epidemic size, the epidemic threshold is the lowest value of $c_p$ for which a nonzero final size can be reached from an infinitesimal initial condition. 
In Fig.~\ref{fig:finalsize_I0}a, we show the final epidemic size, or the stationary number of recovered individuals $R(\infty)$ as a function of the initial number of infected individuals $[I](t=0)$, for various values of the contact tracing rate $c_p$. We consider values of $c_p$ that span the surroundings of the threshold value $c_p^*$ which has been computed analytically using the approximate fast variables method [Eq.~\eqref{eq:threshold_purepairwise}].

Regardless of how small the initial condition is, the final epidemic size is always non-zero, making the numerical threshold blurry. Nevertheless, fitting a line through the final epidemic sizes as functions of the initial conditions, we can estimate the final epidemic size in the limit of a zero initial condition $[I](0)/N \rightarrow 0$. In this limit, we see that the analytical threshold is a good estimate of the numerical threshold, separating a regime where $c_p$ is large and the final epidemic size is zero from a regime where $c_p$ is small and the final epidemic size is non-zero.

\begin{figure}
    \centering
    \includegraphics[width=16cm]{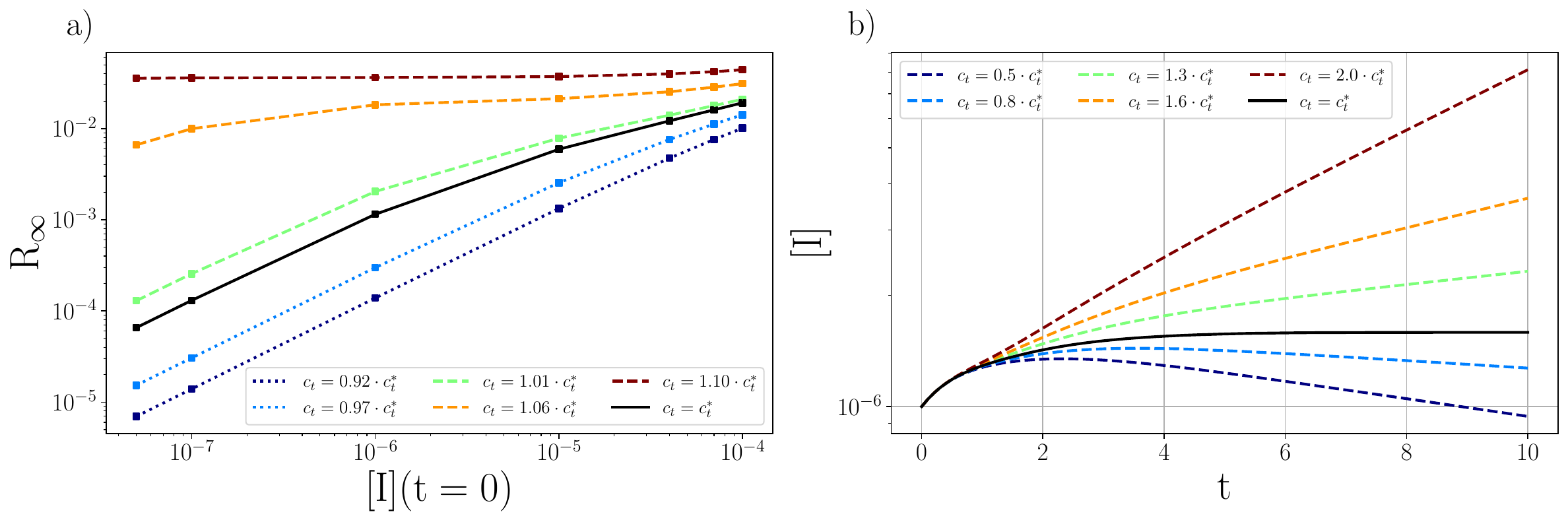}
    \caption{\textbf{Numerical threshold in the pure triplewise contact tracing.} a) The final epidemic size $R(\infty)$ as a function of the initial condition $I(0)$ for different values of the tracing parameter $c_t$. b) The time dependence of the prevalence $[I]$ for different values of the contact tracing rate $c_t$ with initial condition $I(0)=10^{-6}$. Parameter values: $n = 5$, $\tau = 0.18$, $a = 0.6$, $q = 0.6$, $c_t = 0.18$, and $c_p = 0$.}
    \label{fig:finalsize_I0_triple}
\end{figure}

\paragraph{Initial growth rate.}

Concerning the initial growth rate of the epidemic prevalence, the contact tracing parameter $c_p$ plays a slightly delayed role. 
In Fig.~\ref{fig:finalsize_I0}b, we show the time evolution of the prevalence for three values of $c_p$ in the neighborhood of the analytical threshold value $c_p^*$. We find that although the different choices of $c_p$ do not affect the initial growth, after some initial transient time, the contact tracing measure finally kicks in and the role of $c_p$ becomes visible. This transient time happens to also be the time for the fast variables to reach their steady-state. As soon as the fast variables have stabilized, $c_p$ exhibits the following influence: at the threshold value $c_p = c_p^*$ [Eq.~\eqref{eq:threshold_purepairwise}] the prevalence stabilizes to a constant value, when $c_p > c_p^*$ the prevalence is exponentially suppressed, and when $c_p < c_p^*$ it grows exponentially fast. 

The corresponding figures for pure triplewise contact tracing are provided in Fig.~\ref{fig:finalsize_I0_triple}.

\section{Derivation of the critical level of pure pairwise contact tracing}
\label{sec:appendix_derivation}

In this section, we show the main steps of the derivation of the critical level of pure pairwise contact tracing.

Let's consider Eq.~\eqref{eq:tobesolved_explicit}, with $c_t = 0$ (pure pairwise):
\begin{align}
\begin{cases}
\label{system:fast_pairwise_init}
& \tau x - 1 - c_p^* z = 0\\
& \tau x (n-2)-c_p^*\frac{(n-1)}{n} x z - x = 0\\
& 2 \tau x -2 y -2 c_p^* \frac{(n-1)}{n} y z = 0\\
& -a z + c_p^* \left(\frac{(n-1)}{n} y z- \frac{(n-1)}{n} z^2-z\right) + (q y - z) = 0.
\end{cases}
\end{align}

In the second equation, we readily see that $x$ factorizes, while the factor 2 simplifies across the third equation. 
Let $R_0 := \tau (n-2)$ be the basic reproductive number (note that $\tau$ is the ratio between the infection rate and the total recovery rate), and $\kappa := \frac{(n-1)}{n}$. Then Eq.~\eqref{system:fast_pairwise_init} becomes:
\begin{align}
\begin{cases}
\label{system:fast_pairwise}
& \tau x - 1 - c_p^* z = 0\\
& R_0 -c_p^* \kappa z - 1 = 0\\
& \tau x - y - c_p^* \kappa y z = 0\\
& -a z + c_p^* \left(\kappa y z- \kappa z^2-z\right) + q y - z = 0.
\end{cases}
\end{align}

The second equation allows one to express $z$ as a function of the model parameters:
\begin{align}
\label{eq:zz}
    z = \xi \cdot \frac{1}{c_p^*},
\end{align}
where $\xi = \frac{R_0 - 1}{\kappa}$. Note that $z$ depends on the contact tracing parameter $c_p^*$.

Plugging this in the first equation of Eq.~\eqref{system:fast_pairwise}, we can then also express $x$ as a function of the model parameters only:
\begin{align}
\label{eq:xx}
    \tau x = 1 + \xi.
\end{align}

Using the third equation of Eq.~\eqref{system:fast_pairwise}, and plugging the expressions for $x$ and $z$ derived in Eq.~\eqref{eq:xx} and Eq.~\eqref{eq:zz}, we can finally also express $y$ as a function of the model parameters:
\begin{align}
\label{eq:yy}
    y = \frac{1}{R_0} \left(1 + \frac{R_0 - 1}{\kappa} \right) = \frac{1}{R_0} \cdot (1+\xi).
\end{align}

Ultimately, the fourth equation of Eq.~\eqref{system:fast_pairwise} can be used to express the pure pairwise contact tracing rate $c_p^*$ as a function of the fast variables $x$, $y$ and $z$, and the model parameters. Plugging the expressions of the fast variables in terms of the model parameters [Eq.~\eqref{eq:xx}, Eq.~\eqref{eq:yy}, and Eq.~\eqref{eq:zz}], we get:
\begin{align}
    - a \frac{\xi}{c_p^*} + \frac{\kappa}{R_0} (1+\xi) \xi &- \kappa \xi^2 \frac{1}{c_p^*} - \xi + \frac{q}{R_0} (1+\xi) - \xi \frac{1}{c_p^*} = 0\\
    c_p^* &= \frac{R_0 \xi (a+1+\kappa \xi)}{(1+\xi)(\kappa \xi + q) - R_0 \xi}\\
    &= \left(a + R_0\right) \frac{R_0 - 1}{F_p(n, q, \tau)},
\end{align}
where $F_p(n, q, \tau)= \frac{1}{n} \frac{R_0 (nq - 1) + (1-q)}{R_0}$.

\section{Derivation of the critical level of pure triplewise contact tracing}
\label{sec:appendix_derivation_triplewise}

In this section, we show the main steps of the derivation of the critical level of pure triplerwise contact tracing.

Let's consider Eq.~\eqref{eq:tobesolved_explicit}, with $c_p = 0$ (pure triplewise):
\begin{align}
\begin{cases}
\label{system:fast_triplewise_init}
& \tau x - 1 - c_t^* \frac{(n-1)}{n} z^2 = 0\\
& \tau (n-2) - c_t^*\frac{(n-1)(n-2)}{n^2} z^2 - 1 = 0\\
& \tau x - y - c_t^* \frac{(n-1)(n-2)}{n^2} y z^2 = 0\\
& -a z + c_t^*  z^2 \left(\frac{(n-1)(n-2)}{n^2} y - \frac{(n-1)}{n} - \frac{(n-1)(n-2)}{n^2} z\right) + (q y - z) = 0.
\end{cases}
\end{align}

Let $R_0 := \tau (n-2)$ be the basic reproductive number, $\kappa := \frac{(n-1)}{n}$, and $\hat \kappa := \frac{(n-1)(n-2)}{n^2}$. Then Eq.~\eqref{system:fast_triplewise_init} becomes:
\begin{align}
\begin{cases}
\label{system:fast_triplewise}
& \tau x - 1 - c_t^* \kappa z^2 = 0\\
& R_0 - c_t^* \hat \kappa z^2 - 1 = 0\\
& \tau x - y - c_t^* \hat \kappa y z^2 = 0\\
& -a z + c_t^* z^2 \left(\hat \kappa y - \kappa - \hat \kappa z\right) + q y - z = 0.
\end{cases}
\end{align}

The second equation allows one to express $z^2$ as a function of the model parameters:
\begin{align}
\label{eq:zz_t}
    z^2 = \frac{\xi}{\kappa} \cdot \frac{1}{c_t^*},
\end{align}
where $\xi = \kappa \frac{R_0 - 1}{\hat \kappa}$. Note that $z^2$ depends on the contact tracing parameter $c_t^*$.

Plugging this in the first equation of Eq.~\eqref{system:fast_triplewise}, we can then also express $x$ as a function of the model parameters only:
\begin{align}
\label{eq:xx_t}
    \tau x = 1 + \xi.
\end{align}

Using the third equation of Eq.~\eqref{system:fast_pairwise}, and plugging the expressions for $x$ and $z^2$ derived in Eq.~\eqref{eq:xx_t} and Eq.~\eqref{eq:zz_t}, we can finally also express $y$ as a function of the model parameters:
\begin{align}
\label{eq:yy_t}
    y = \frac{1}{R_0} \cdot (1+\xi).
\end{align}

Ultimately, the fourth equation of Eq.~\eqref{system:fast_triplewise} can be used to express the pure triplewise contact tracing rate $c_t^*$ as a function of the fast variables $x$, $y$ and $z$, and the model parameters. Plugging the expressions of the fast variables in terms of the model parameters [Eq.~\eqref{eq:xx_t}, Eq.~\eqref{eq:yy_t}, and Eq.~\eqref{eq:zz_t}], we get:
\begin{align}
    - z (a + 1 + R_0 &- 1) + q \frac{1 + \xi}{R_0} + \frac{\xi}{\kappa} (\hat \kappa \frac{1+\xi}{R_0} - \kappa) = 0\\
    \frac{1}{z} &= \frac{(a + R_0)}{q \frac{1 + \xi}{R_0} + \frac{R_0-1}{R_0}(1+\xi) - \xi}\\
    c_t^* = \frac{(a+R_0)^2}{(q \frac{1 + \xi}{R_0} + \frac{R_0-1}{R_0}(1+\xi) - \xi)^2} \frac{\xi}{\kappa}&
    = \frac{(a+R_0)^2 (R_0-1)}{\hat \kappa (q \frac{1 + \xi}{R_0} + \frac{R_0-1}{R_0}(1+\xi) - \xi)^2}
     = \left(a + R_0\right)^2 \frac{R_0 - 1}{F_t(n, q, \tau)},
\end{align}
where $F_t(n, q, \tau) = \frac{(n-1)}{n^2(n-2)} \left(\frac{R_0 (n q - 2) + 2(1-q)}{R_0}\right)^2$. 

\end{document}